\documentclass[12pt,preprint]{aastex}
\begin {document}

\title{The Asymmetrical Wind of the Candidate Luminous Blue
Variable MWC 314}

\author{John P. Wisniewski\altaffilmark{1}, Brian L. 
Babler\altaffilmark{2},
Karen S. Bjorkman\altaffilmark{3}, Anatoly V. 
Kurchakov\altaffilmark{4}, Marilyn R. Meade\altaffilmark{2}, 
Anatoly S. Miroshnichenko\altaffilmark{5}}

\altaffiltext{1}{Universities Space Research
Association/NASA GSFC Code 667, Building 21, Greenbelt, MD 
20771, jwisnie@milkyway.gsfc.nasa.gov}
\altaffiltext{2}{Space Astronomy Lab, University of Wisconsin-Madison,
1150 University Avenue, Madison, WI 53706} 
\altaffiltext{3}{Ritter Observatory, MS \#113, Department of Physics
and Astronomy, University of Toledo, Toledo, OH 43606}
\altaffiltext{4}{Fesenkov Astrophysical Institute, 050020, Almaty Observatory,
Kazakhstan}
\altaffiltext{5}{Department of Physics \& Astronomy, 
University of North Carolina-Greensboro, P.O. Box 26170, Greensboro, NC
27402}

\begin{abstract}

We present the results of long-term spectropolarimetric and spectroscopic monitoring
of MWC 314, a candidate Luminous Blue Variable star.  We detect the first
evidence of H$\alpha$ variability in MWC 314, and find no apparent
periodicity in this emission.  The total R-band polarization is 
observed to vary between 2.21\% and 3.00\% at a position angle consistently
around $\sim$0$^{\circ}$, indicating the presence of a time-variable
intrinsic polarization component, hence an asymmetrical circumstellar
envelope.  We find suggestive evidence that MWC 314's intrinsic 
polarization exhibits a wavelength-independent magnitude varying
between 0.09\% and 0.58\% at a wavelength-independent position angle
covering all four quadrants of the Stokes Q-U plane.  Electron
scattering off of density clumps in MWC 314's wind
is considered as the probable mechanism responsible for these
variations.

\end{abstract}


\section{Introduction}

As massive stars progress through the post-main-sequence evolutionary phase of 
their lifetimes, at least some will briefly enter the Luminous Blue
Variable (LBV) evolutionary stage.  Comprehensive reviews of LBVs 
and the properties of their circumstellar environments are
given by \citet{hum94} and \citet{van01}.  Summarized briefly, LBVs 
are extremely luminous stars which exhibit photometric variability at
multiple levels over a multitude of timescales, including large 
eruptions of up to 5 magnitudes on timescales of years to centuries,
complex spectroscopic variability, and evidence of visible ejecta.

Only a limited number of LBVs or candidate LBVs (cLBVs) 
have been identified to date (e.g., 
\citealt{van01}), hindering efforts to understand the underlying
physics which drive the variability of these objects.  
\citet{mir96} 
analyzed optical and near-IR photometry as well as optical spectroscopy
of MWC 314 and concluded it was one of the most luminous 
stars in the Milky Way.  While there is no record of MWC 314 undergoing
a large photometric eruption and it doesn't show evidence of a dust
shell, \citet{mir96} found MWC 314's fundamental
parameters were similar to those of LBVs and thus suggested the object be
classified as a cLBV.  Followup high-resolution spectroscopy 
\citep{mir98} revealed MWC 314 had a N/O overabundance similar to that
of other known LBVs; however, no evidence of spectral line variability
was detected.  These authors also suggested that MWC 314 possessed
a non-spherical wind and that its circumstellar envelope was not oriented
edge-on.

Polarimetry is well established as an effective tool to probe the
geometry of circumstellar environments.  Polarimetry has been utilized
to investigate the environments of the LBVs AG Car \citep{lei94,sch94,dav05},
HR Car \citep{cla95, par00,dav05}, R127 \citep{sch93}, $\eta$ Car \citep{sch99,dav05},
P Cyg \citep{hay85,lup87,tay91,nor01,dav05}, and numerous SMC/LMC LBVS \citep{dav05}.  
\citet{sch93} suggested a
link between the geometry of the present-day wind in R127, diagnosed via
polarimetric observations, and the geometry of an older eruption which
created a shell.  Linking previous large mass-loss events with current
stellar activity can provide important constraints for efforts to 
understand the physics behind massive LBV eruptions.

We have been monitoring the candidate LBV MWC 314
spectropolarimetrically for 6 years and spectroscopically for 4 years 
in an effort to diagnose its 
circumstellar environment and compare its properties with those of
known LBVs.  In Section 2, we summarize the
acquisition and reduction of these data.  The spectroscopic variability of
MWC 314 is detailed in Section 3.1 and the variability of its total polarization
is outlined in Section 3.2.  In Section
3.3, we discuss our efforts to identify the interstellar polarization
along the line of sight to MWC 314, while Section 3.4 describes the
intrinsic polarization component which results after the interstellar
component has been removed.  A discussion of the implications of these
data is presented in Section 4.

\section{Observations}

MWC 314 was monitored from August 1999 through September 2004 with
the HPOL spectropolarimeter, mounted on the 0.9m telescope at the 
University of Wisconsin's Pine Bluff Observatory (PBO).  Sixteen observations were
made using the red grating, providing spectropolarimetric coverage from 
$\sim$6000-10500 \AA\ at a resolution of $\sim$10 \AA\ \citep{nor96}, and
were recorded with a 400 x 1200 pixel CCD camera.  MWC 314 was observed
using a 6 
arc-second slit aligned E-W and a 200 arc-second decker aligned N-S on
the sky. 
Data were processed using UW's REDUCE software package\citep{wol96}: 
further information regarding HPOL and REDUCE are detailed in 
\citet{noo90} and \citet{har00}.

The instrumental polarization of HPOL has been monitored on an approximately monthly
basis since HPOL's inception in 1989 via observations of polarized
and unpolarized standard stars.  Our data have been corrected for such
instrumental effects to an absolute accuracy of $0.025\%$ and
1$^{\circ}$ in the R band.  As
noted in \citet{har00}, HPOL spectroscopic data are not calibrated to
an absolute flux level due to the variable sky conditions routinely
present.

Additional spectroscopic observations of MWC 314 were obtained with the
1-m RCC telescope of the Assy-Turgen high-altitude Observatory of the
Fesenkov Astrophysical Institute (FAI) (Kazakhstan).  Spectra 
covering the wavelength regime 6500-6700 \AA\ were acquired
with a standard UAGS spectrograph and a 1530 x 1020 ST-8Ei CCD camera
with pixels 9 x 9 mm in size.  The reciprocal dispersion of the 
instrument is 0.5 \AA\ mm$^{-1}$. 
A 3 arc-second slit (3-pixel wide) was used for the observations.
For flat-fielding, reflection spectra of the telescope dome illuminated by
an ordinary tungsten-filament light bulb were recorded. The spectra were not
corrected for the instrumental profile function. The H$\alpha$ equivalent
widths (EWs) were measured with no correction for the atmospheric absorption
spectrum.

\section{Results}
\subsection{Spectroscopy}

Table 1 summarizes the observational data which are analyzed in this
study.  \citet{mir98} used high-resolution spectroscopy to document
the wealth of optical emission lines which MWC 314 exhibits; however, they
did
not detect evidence of variability in their limited dataset.  To search
for evidence of H$\alpha$ variability, we measured the H$\alpha$ EWs 
of our dataset.  For our spectropolarimetric dataset, the 
spectral regions spanning
6300-6400 \AA\ and 6720-6820 \AA\ were used to measure the
continuum flux level around H$\alpha$, while the
region $\sim$6518-6622 \AA\ was used to determine the H$\alpha$ line flux.
A limited number of our observations suffered from less than perfect
wavelength calibration: we took great care to ensure our measurements
of line flux always probed the same spectral region in every
observation.  Our FAI spectra covered the region 6500-6700 \AA.  Data
were continuum normalized using line-free regions in this spectral
range, and EWs were systematically determined using
standard IRAF techniques.  

We tabulate our EW measurements in Table 1 and plot their
time dependence in Figure 1.  Our FAI EWs appear to be systematically larger than those derived from 
our PBO spectropolarimetric dataset by $\sim$28 \AA.  We believe that the wings of the H$\alpha$ emission 
in MWC 314 are not fully contained within the spectral bandpass of the FAI data, leading to improper 
continuum normalization and hence the systematic inflation of measured EWs.  We have recorded the 
actual measured EWs for our FAI data in Table 1, and have applied a -28 \AA\ shift to these data 
in Figure 1.  Errors associated with our equivalent
width measurements are dominated by continuum placement uncertainty, which we estimate contributes 
an uncertainty of $<$ 5\% to all measurements; photon statistics errors are less than 1 \AA.

Inspection of Table 1 and Figure 1 shows that
the H$\alpha$ EW often is seen at a level of 142 \AA, consistent with
that measured by \citet{mir96} and \citet{mir98}.  Contrasting these
earlier observations, our data show distinct 
deviations from this average EW, with emission strengths varying from 115 \AA\
to 149 \AA.  We observe no periodicity in this changing emission
strength, although our sparse sampling likely masks the presence of any
true periodic behavior inherent in our data.  Interestingly, our FAI
spectra, which were obtained at a higher sampling frequency than our PBO
data, exhibits substantial variability on timescales of several days to
weeks.  We encourage future moderate resolution spectroscopic monitoring
with a higher sampling frequency to search for both evidence of binarity and
detailed line profile variability.

\subsection{Total Polarization Variability}

It is well established that the polarized light we observe from
astronomical sources can be comprised of two distinct components: 
an intrinsic polarization component originating at a source or its immediate
vicinity and an interstellar polarization (ISP) component created by the
dichroic absorption of starlight along a line of sight by aligned dust
grains.  For unresolved sources such as MWC 314, an intrinsic
polarization component will be observed if either the source has an
asymmetrical circumstellar geometry or the illumination source of
a symmetrical circumstellar envelope is time-dependent \citep{bjo00}.  
Conversely, one
would not expect the ISP along a line of sight to vary; hence, observing
variability in the total polarization of an object is equivalent to
detecting the presence of an intrinsic polarization component.

To better visualize any polarimetric variability present in our data,
we have binned, i.e. reduced the resolution of our data to 
a standard Johnson R filter, as seen in Figure 1 and tabulated in 
Table 1.  These broadband polarization measurements clearly exhibit
a significant time-dependent behavior, with R-band polarizations
ranging from 2.21\% to 3.00\% at position angles consistently around
0$^{\circ}$.  Thus MWC 314 must possess a variable intrinsic
polarization component.  There does not appear to be any firm
correlation between the time variability of the total R band
polarization and the H$\alpha$ EW.

Figures 2 presents the total polarization of each of our
observations, after the data was binned to a constant error level.  Although not
pictured here, the polarization position angle of these data exhibited a near constant, 
wavelength independent value of $\sim$179$^{\circ}$.  A dashed vertical line 
through these Figures denotes the wavelength of
H$\alpha$.  Note that quite a few of our observations show distinct
polarization features at H$\alpha$, namely depolarization effects.  
In Section 3.3 we will discuss how this depolarization phenomenon
can be used as a tool to constrain the ISP component along the line of sight. 

\subsection{Interstellar Polarization}

Because polarization components add vectorally, it is possible to
isolate and study the behavior of intrinsic polarization signatures of
stars if their ISP components are accurately
identified.  Numerous techniques have been established for
removing ISP, as detailed by \citet{mcl79}, \citet{qui97}, and \citet{bjo06}. 
Each technique has its limitations, some of which we now
review.

A traditional way of determining foreground ISP is via field star
studies.  If one can locate a sample of stars which a) are within
a small angular separation from and at the same general distance to the target 
object, such that every star's light will pass through the same
interstellar medium conditions and b) exhibit no
intrinsic polarization, then the averaged polarization of these field stars
provides an estimate of the ISP along the particular line of sight. 
\citet{mir98} determined a distance to MWC 314 of 3.0 $\pm$ 0.2 kpc:
searches of polarization catalogs such as \citet{hei00} do not show
any suitable field stars which meet the aforementioned requirements.
HD 174571 is a small angular distance away from MWC 314 and we estimate 
its distance to be $\sim$0.5 kpc based upon its brightness and reddening 
\citep{vie03}; however, it is classified as a Be star and thus likely 
has a variable intrinsic polarization component.  HD 183143, a 
suggested polarization standard star which has been observed to be
variable at a level of 0.3\% \citep{hsu82}, is also located at a small
angular distance from MWC 314.  Based upon its distance, $\sim$1.6 kpc
from its brightness and reddening, HD 183143 shows an enormous
interstellar polarization of $\sim$6.0\% at a PA of $\sim$0$^{\circ}$ 
in V, which \citet{sch83} notes differs greatly from the typical ISP
value of 1-3\% at a PA of $\sim$60$^{\circ}$ found in that region 
\citep{mat70}.  It thus seems likely that the interstellar medium near
MWC 314's line of sight is quite complex, implying that the field star
technique is not an appropriate tool for determining MWC 314's ISP.

Another technique often employed to determine ISP
components is to measure the polarization across intrinsically
unpolarized emission lines.  Such a
technique makes the assumption that most recombination events occur outside of
the innermost circumstellar region which polarizes light \citep{har68}. 
 Polarization
can simply be viewed as the ratio of scattered to total (scattered plus
unscattered) intensity.  The influx of a large number of unpolarized H$\alpha$ 
line photons will significantly dilute the presence of any polarized continuum
photons, causing the emission line to appear ``depolarized''.
Thus the polarization measured across emission lines should be
purely interstellar in origin.  Such an assumption is not always
fully accurate \citep{mcl79,qui97}; however, the technique has been used
to provide ISP estimates in previous studies of LBVs 
\citep{tay91,sch93,sch94,nor01}.  Inspection of Figure 2 indeed shows
that most of our observations exhibit H$\alpha$ depolarization effects.
We measured the H$\alpha$ and continuum
polarization for each of our observations, using the continuum bandpass 
definition discussed in Section 3.1 and a 50 \AA\ bandpass, centered on
H$\alpha$, to sample the core of the H$\alpha$ line.  The
results are plotted on a Stokes Q-U diagram in Figure 3, where solid
symbols correspond to line polarizations, open symbols represent
continuum polarizations, and error bars correspond to 1-$\sigma$ 
photon statistics errors.  Solid lines connect the line and continuum
polarization for each observation.  Most of these lines 
point to one specific area in the Q-U diagram, i.e. the
likely ISP for MWC 314.  With the exception of the 4 vectors which
do not point to this common area, specifically data from 6 August 1999, 
8 August 2002, 3 September 2003, and 3 June 2004, we averaged all line
polarizations together using a 1/$\sigma^{2}$ weighting technique.
The resulting Stokes parameters, Q = 2.59\% and U = -0.14\% (P = 
2.59\% PA = 179$^{\circ}$), define our estimate of the ISP at 6563 \AA.
In Section 4, we will further discuss the implications of excluding the
4 aforementioned nights in this determination of MWC 314's ISP.

We can parameterize the ISP using the empirical Serkowski law\citep{ser75},
as modified by \citet{wil82}, \begin{equation} P_{\lambda} = 
P_{max} \times exp[-K \times ln^{2}(\lambda_{max} / \lambda)], 
\end{equation} where P$_{max}$ is the maximum polarization and
$\lambda_{max}$ is the wavelength at which P$_{max}$ occurs.  Assuming
a typical value of $\lambda_{max}$, 5500 \AA, and our measured ISP at
H$\alpha$, we determine P$_{max}$ to be 2.67\%.  In summary, our ISP  
parameters are P$_{max}$ = 2.67\%, $\lambda_{max}$ = 5500 \AA,  
$\theta$ = 179$^{\circ}$, and K = 0.923.

\subsection{Intrinsic Polarization Variability}

Using the ISP parameters discussed in Section 3.3, we removed the
interstellar polarization from our data.  To check the accuracy of
this process, we measured the intrinsic polarization of H$\alpha$ and
its surrounding continuum for our data using the same bandpass definitions
discussed in Section 3.3.  We plot the results on a Stokes Q-U diagram,
seen in Figure 4, where closed symbols again represent H$\alpha$
line polarization and open symbols represent continuum polarization 
measurements.
With the exception of the 4 previously identified anomalous
observations, all H$\alpha$ line polarizations congregate around the
origin as expected if these lines exhibit only ISP.
  Interestingly, the line-to-continuum polarization data 
trace out radial patterns from the origin in random directions,
indicating there exists no consistent position angle in the 
intrinsic polarization. 

Figures 5 and 6 illustrate the wavelength dependence of the
intrinsic polarization of our data.  The magnitude of the intrinsic
polarization is clearly time dependent, with R-band polarizations 
varying from 0.09\% to 0.58\%.  The intrinsic continuum
polarization and position angle for each night are both generally wavelength 
independent, a signature of polarization arising from electron
scattering with no subsequent attenuation.  The polarization position angles fill all four quadrants
of Stokes Q-U space, indicating there is no single scattering 
geometry present.  Note that whenever polarization measurements approach
zero their corresponding position angles are poorly
constrained; hence, the large error bars present in some data plotted
in Figure 5 and 6 should not be construed as excessively noisy data.

\section{Discussion}

Recall that we used the presence of depolarized emission lines in many
of our observations to isolate and remove the ISP
along the line of sight to MWC 314.
In the process of determining this ISP component, we
remarked that several observations showed abnormal H$\alpha$ 
line polarizations
(e.g., Figure 3) and omitted these data from our ISP determinations.
Using the polarization of emission lines as a basis for determining ISP
parameters assumes that most recombination
events occur outside of the region immediately surrounding a star which
has a high free electron density, i.e. the polarizing region.
If many H$\alpha$ recombination photons do experience a scattering
event in the polarizing region, then intrinsic H$\alpha$ line 
polarization can arise.
We suggest that the sporadic observations which show intrinsic H$\alpha$
line polarization, specifically the 6 August 1999 (Figure 5) and 8
August 2002 (Figure 6) data, arise from density 
enhancements near the base of MWC 314's wind.  We would expect that as 
these density
clumps propagate away from and exit the polarizing region of the wind,
the intrinsic polarization of H$\alpha$ would decrease to its nominal
``depolarized'' state, e.g. transition to exhibiting zero intrinsic
polarization.  

The geometry of MWC 314's circumstellar envelope was previously
explored by \citet{mir98}, who suggested the envelope was non-spherical
and not viewed edge-on, based upon Balmer line profiles.
Our data clearly illustrate that MWC 314 exhibits a time-dependent
intrinsic polarization component, indicating the presence of an
asymmetrical circumstellar envelope.  The wavelength-independent 
nature of MWC 314's intrinsic polarization strongly suggests that
electron scattering is the dominant polarizing mechanism.  The
position angle of this intrinsic polarization varies throughout 
all four quadrants of the Stokes Q-U plane, indicating that there
is no preferred scattering plane, such a disk, in MWC 314's envelope.
We note that the LBV P Cygni exhibits an intrinsic polarization whose
position angle also varies significantly over time, which  
\citet{tay91} and \citet{nor01} interpreted as evidence of
electron scattering in a clumpy wind (see Figure 10a of \citealt{tay91} for a graphical 
illustration).  We have previously speculated 
that the sporadic observations of
intrinsic H$\alpha$ line polarization may be due to clumps near the
base of MWC 314's wind; we also suggest that, like P Cygni, the 
general behavior of MWC 314's
intrinsic polarization and position angle may be explained by
electron scattering in an envelope characterized by density 
inhomogeneities, e.g. clumps.  As the distribution and density of
these clumps in the wind change over time, the globally averaged
polarization position angle and magnitude will also fluctuate
over the entire Q-U plane.  

As detailed by \citet{not95}, the majority of LBVs exhibit some type
of axisymmetric circumstellar geometry which likely influence the
morphology of their ejected nebulae.  \citet{sch93}, for example, 
noted that the morphology of the LBV R137's nebula seemed to correspond
to the circumstellar geometry derived from their polarimetric
observations.  The major exception to this basic geometry is P Cygni,
which seems to exhibit a spherically symmetric nebula characterized
by an asymmetrical distribution of clumps
\citep{tay91,bar94,ski97}.  
Our observations of MWC 314 do not reveal
the presence of an axisymmetric circumstellar environment; rather, 
its environment is more similar to that observed for P Cygni,
i.e. an asymmetrical clumpy wind.  
We thus speculate that, like P Cygni, if MWC 314 undergoes an 
eruptive outburst in the near future, its morphology will be
generally spherically symmetric and characterized by an asymmetrical 
distribution of density enhancements.

We suggest a number of followup observations which would further 
refine our understanding of MWC 314's circumstellar environment.
Our spectroscopic observations showed clear signs of H$\alpha$ 
variability, yet much better data sampling will be required to
search for the short- and long-term periodic behavior common amongst many
LBVs.  Detailed information regarding the nature of the clumpy
wind we detected can be derived from our data (e.g., 
\citealt{cas87, tay91, nor01}) if we can refine the inclination
of MWC 314's extended envelope.  Searches for remnants of 
past eruption events, which
may still exhibit diffuse radio or infrared emission, should also
be attempted.

\section{Summary}

We have presented the results from a longterm spectroscopic and spectropolarimetric 
study of the candidate
Luminous Blue Variable MWC 314.  We have detected the first conclusive
evidence of H$\alpha$ equivalent width variability.  The observed
variability of MWC 314's total polarization indicates the presence
of an intrinsic component, hence an asymmetrical circumstellar
environment.  Using depolarized H$\alpha$ emission, we determined 
the ISP along the line of sight to MWC 314,
allowing us to study its intrinsic component in detail.  We find
MWC 314's intrinsic polarization is wavelength independent and
exhibits no single scattering geometry.  We suggest that electron 
scattering off of a time variable distribution of density clumps
in the wind is responsible for the behavior of this intrinsic polarization
component.  We find occasional evidence of intrinsic H$\alpha$ line
polarization and suggest such events are probing density enhancements
at the base of MWC 314's wind.  Finally, we note similarities between
the polarimetric properties of MWC 314 and the LBV P Cygni, and
suggest that any future eruptive outburst from MWC 314 may resemble
the morphology exhibited by P Cygni.

\acknowledgments

We would like to thank Ken Nordsieck for providing access to
the HPOL spectropolarimeter, and the PBO observing team for obtaining these
data.  We also thank the annonymous referee for providing helpful suggestions to 
improve this paper.  This work has been supported in part by NASA LTSA grant NAG5-8054
(KSB, ASM) and NASA GSRP fellowship NGT5-50469 (JPW).  K.S.B. is a Cottrell 
Scholar of the Research Corporation, and gratefully acknowledges their
support.  Partial support for HPOL has been provided by NASA grant
NAS5-26777.  This research has made use of the SIMBAD database operated at
CDS, Strasbourg, France, and the NASA ADS system.

\clearpage
\newpage
\begin{figure}
\includegraphics[scale=1.0]{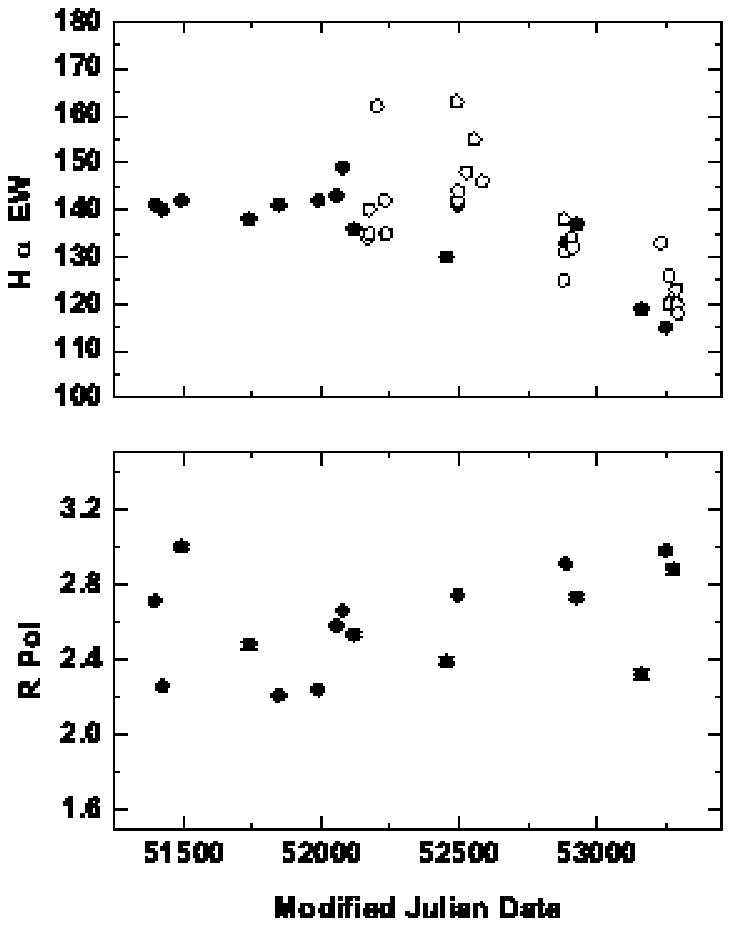}
\figcaption[f1.eps]{H$\alpha$ equivalent widths and R band polarizations
of MWC 314 are plotted as a function of time.  Open circles correspond to FAI spectroscopic observations, which as 
discussed in the text have been shifted in magnitude by -28 \AA.
  Filled circles correspond to HPOL spectropolarimetric data.  The dominant source of error in these 
measurements is our choice of continuum placement; the photon statistic errors associated with 
these measurements are less than 1 \AA. Note that the polarization error bars are often smaller than
the size of the data points.}
\end{figure}

\clearpage
\newpage
\begin{figure}
\includegraphics[width=5.0in,angle=0]{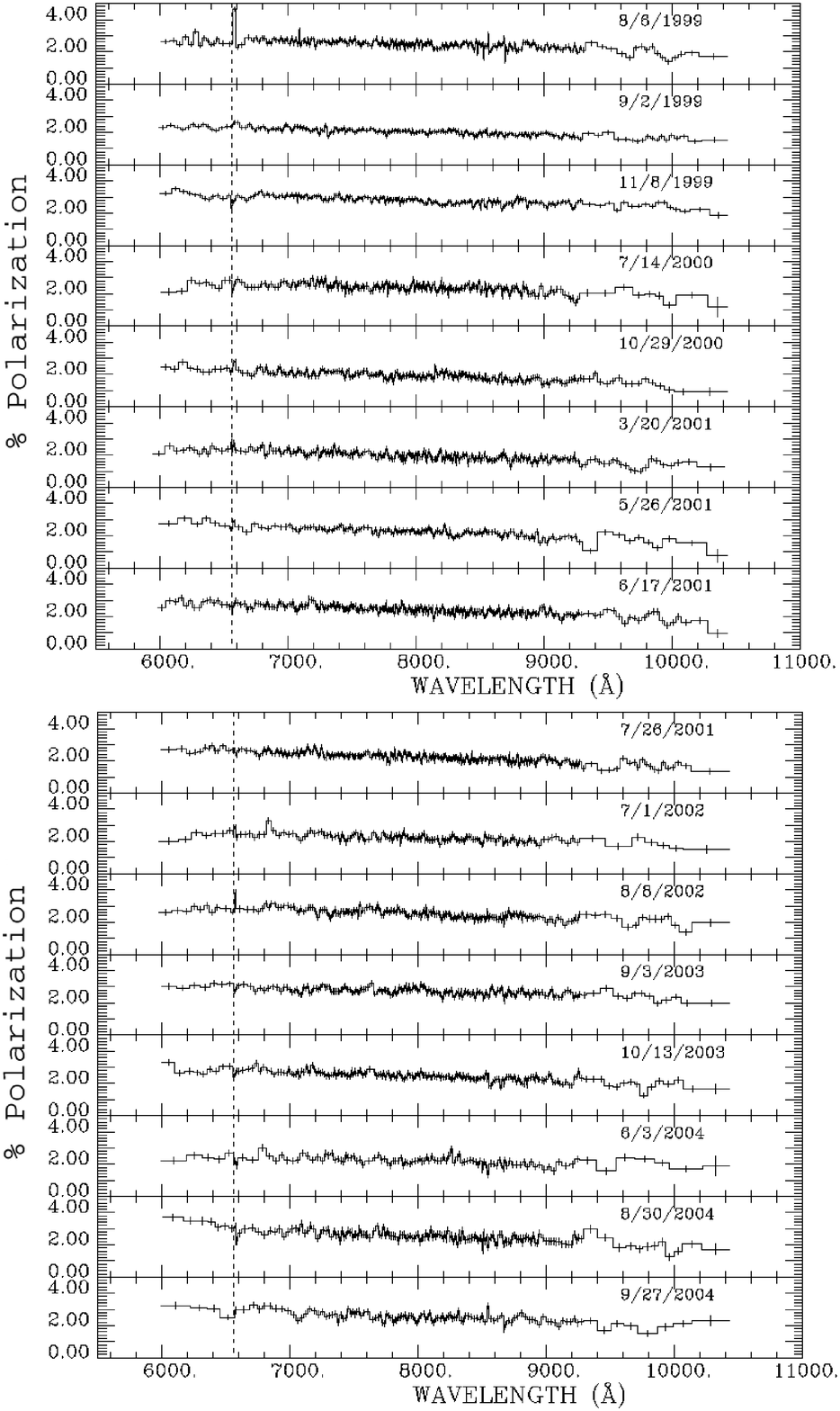}
\figcaption[f2.eps]{Time variability of MWC 314's total R band polarization.  The dashed 
vertical line indicates the position of H$\alpha$.}
\end{figure}

\clearpage
\newpage
\begin{figure}
\center
\includegraphics[width=4.0in]{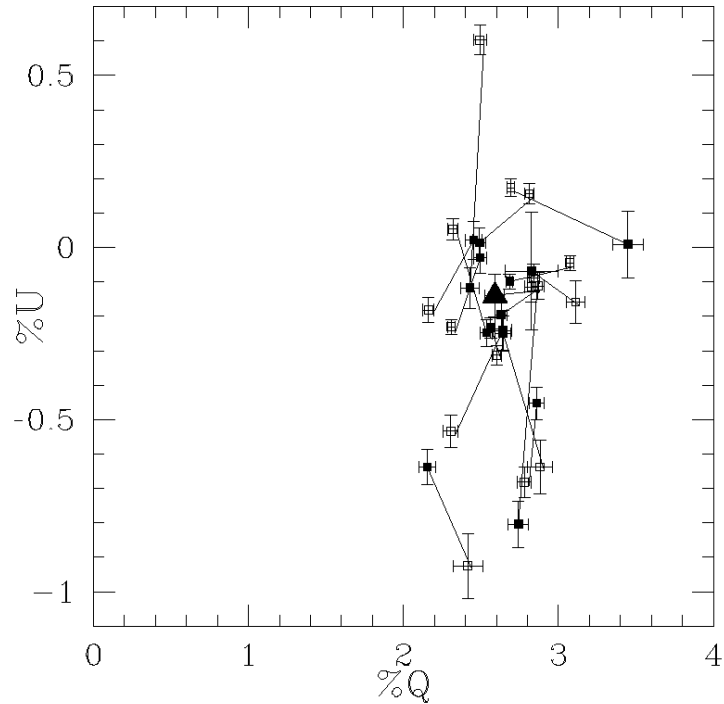}
\figcaption[f3.eps]{Q-U diagram of MWC 314's total polarization.  Solid
squares denote the polarization at H$\alpha$ while open squares denote
the continuum polarization.  Note that most of the lines connecting an
observation's continuum to H$\alpha$ polarization point to one location
in the Q-U diagram.  Assuming H$\alpha$ should generally be unpolarized,
this location identifies the ISP along the line
of sight.  The ISP we calculated by averaging most of these H$\alpha$
polarizations is given by the large, solid triangle, located at Q = 2.83\% and U = -0.07\%.}
\end{figure}

\clearpage
\newpage
\begin{figure}
\includegraphics[width=4.0in]{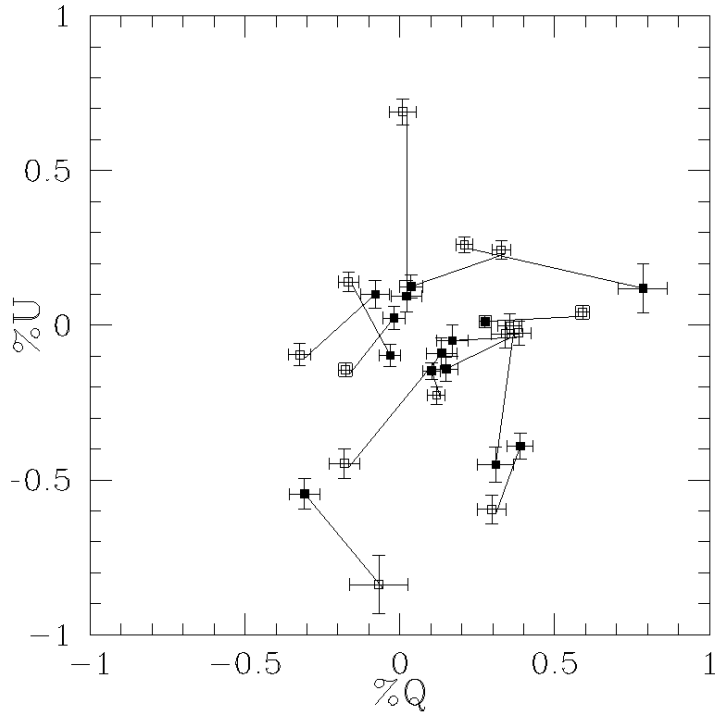}
\figcaption[f4.eps]{Q-U diagram of MWC 314's intrinsic polarization.
Note that most
of the H$\alpha$ line polarizations, to within 3-$\sigma$, now fall near the
origin.  Such a scenario is expected if H$\alpha$ is generally
unpolarized.  The lines connecting each observation's line to continuum
polarization are oriented randomly about the origin, indicating that
there is no single preferred position angle to the intrinsic
polarization component.}
\end{figure}

\clearpage
\newpage
\begin{figure}
\includegraphics[width=5.0in,angle=0]{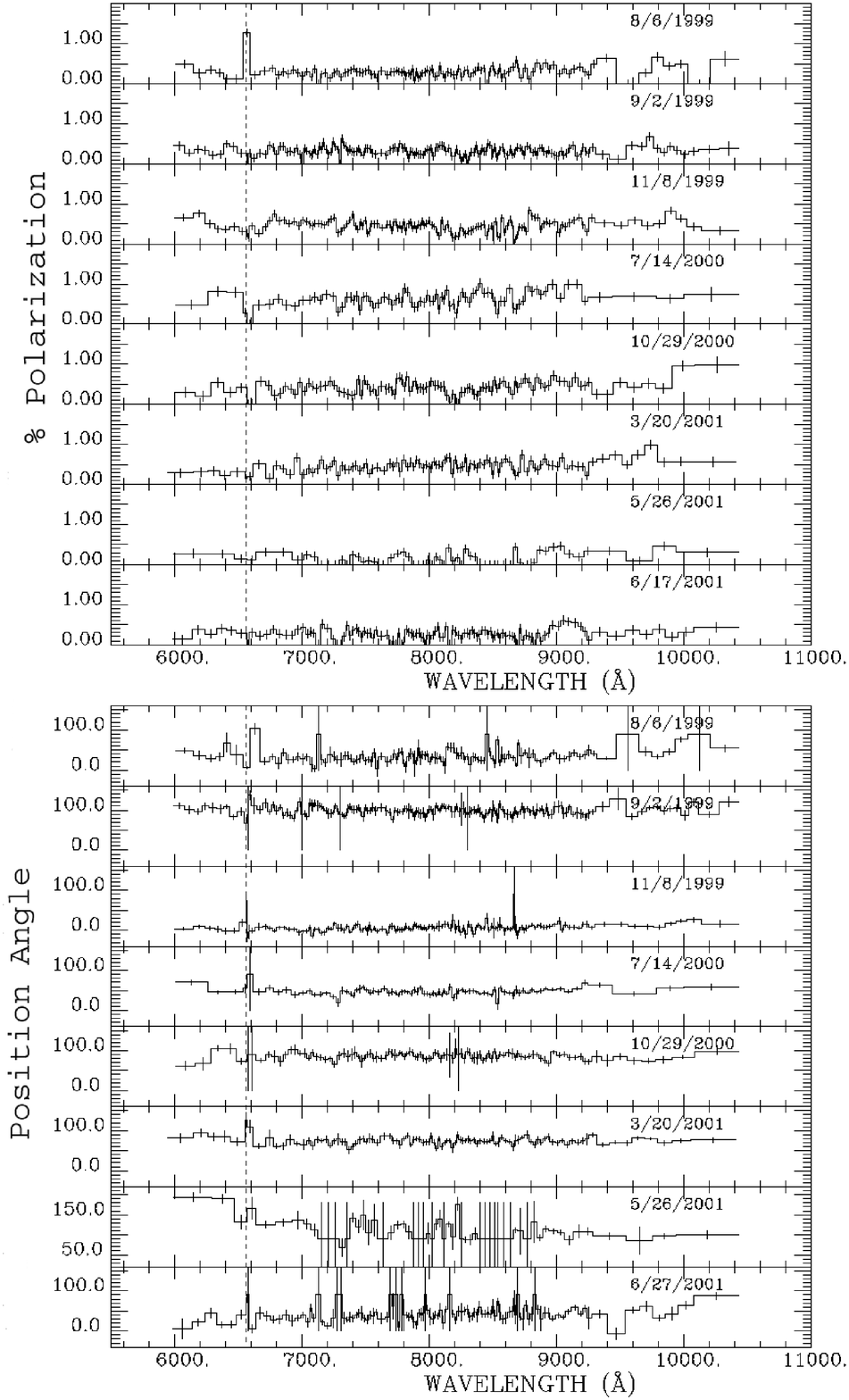}
\figcaption[f5.eps]{Time variability of MWC 314's total R band intrinsic 
polarization and position angle for the first half of our dataset.  The dashed vertical line indicates the 
position of H$\alpha$.  Note that the polarization and position angle for each observation
are generally wavelength-independent, indicating electron scattering
is the dominant polarizing mechanism.}
\end{figure}

\clearpage
\newpage
\begin{figure}
\includegraphics[width=5.0in,angle=0]{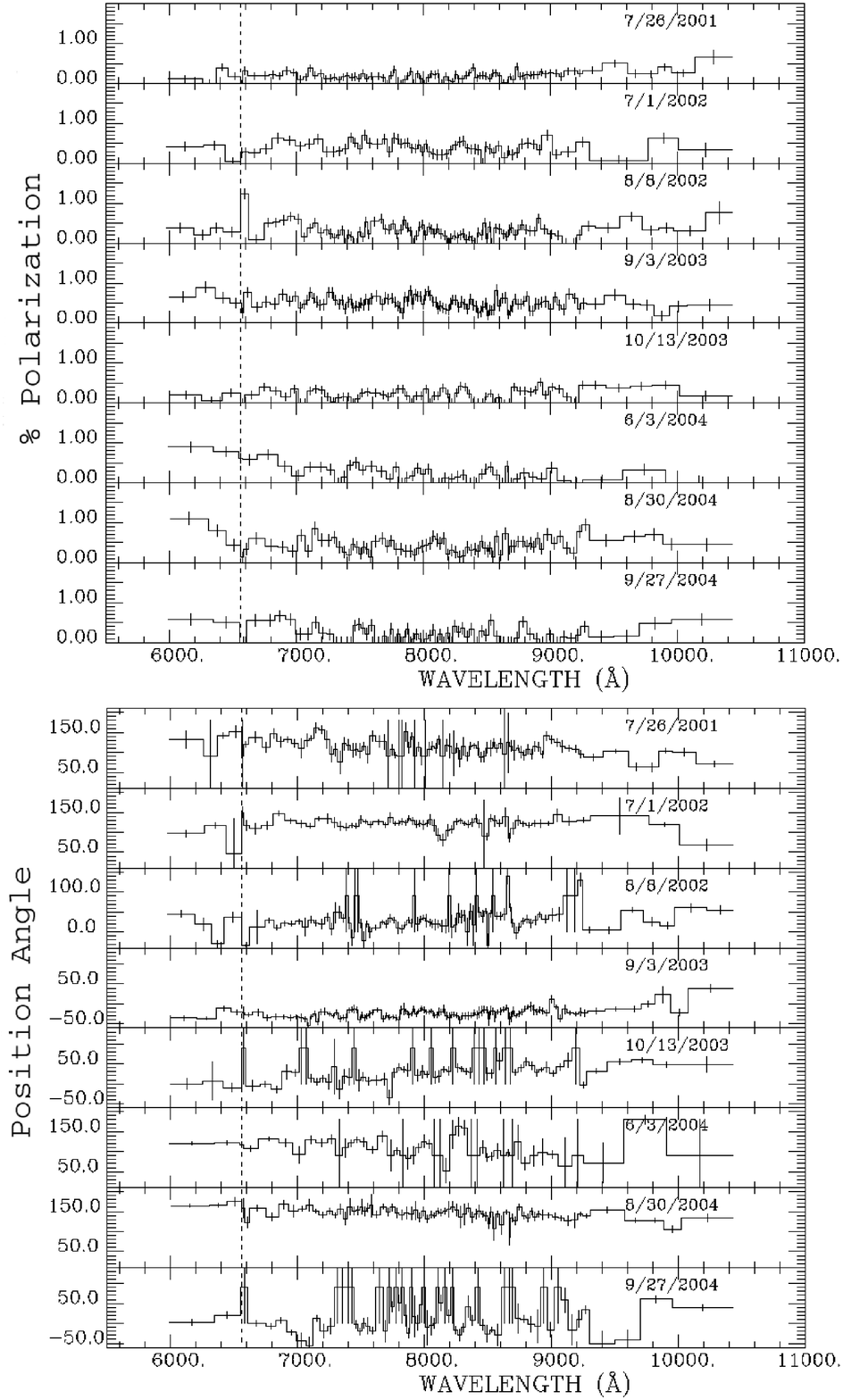}
\figcaption[f6.eps]{Time variability of MWC 314's total R band intrinsic
polarization and position angle for the second half of our dataset.  The dashed vertical line indicates the 
position of H$\alpha$.  Note that the polarization and position angle for each observation
are generally wavelength-independent, indicating electron scattering
is the dominant polarizing mechanism.}
\end{figure}

\clearpage
\newpage
\begin{table}
\caption{}
\scriptsize
\begin{tabular}{lccccc}
Date & MJD & Location & H$\alpha$ EW & \%Pol$_{R}$ & PA$_{R}$ \\
\tableline

8-6-1999 &   51397 & P & 141  & 2.71 $\pm$ 0.01\% & 1 \\
9-2-1999 &   51424 & P & 140  & 2.26 $\pm$ 0.01\% & 178 \\
11-8-1999 &  51491 & P & 142  & 3.00 $\pm$ 0.01\% & 0 \\
7-14-2000 &  51740 & P & 138  & 2.48 $\pm$ 0.02\% & 5 \\
10-29-2000 & 51847 & P & 141  & 2.21 $\pm$ 0.01\% & 0 \\
3-20-2001 &  51989 & P & 142  & 2.24 $\pm$ 0.01\% & 1 \\
5-26-2001 &  52056 & P & 143  & 2.58 $\pm$ 0.01\% & 178 \\
6-17-2001 &  52078 & P & 149  & 2.66 $\pm$ 0.01\% & 1 \\
7-26-2001 &  52117 & P & 136  & 2.53 $\pm$ 0.01\% & 177 \\
9-16-2001 & 52169 & F & 162  & \nodata & \nodata \\
9-20-2001 & 52173 & F & 163 & \nodata & \nodata \\ 
9-21-2001 & 52174 & F & 168 & \nodata & \nodata \\  
10-18-2001 & 52201 & F & 190 & \nodata & \nodata \\  
11-16-2001 & 52230 & F & 170 & \nodata & \nodata \\   
11-18-2001 & 52232 & F & 163 & \nodata & \nodata \\  
11-21-2001 & 52235 & F & 163 & \nodata & \nodata \\ 
7-1-2002 &   52457 & P & 130  & 2.39 $\pm$ 0.02\% & 175 \\
8-5-2002 & 52492 & F & 191 & \nodata & \nodata \\   
8-6-2002 & 52493 & F & 170 & \nodata & \nodata \\  
8-8-2002 &   52495 & P & 141  & 2.74 $\pm$ 0.01\% & 0 \\
8-9-2002 & 52496 & F & 172 & \nodata & \nodata \\ 
9-7-2002 & 52525 & F & 176 & \nodata & \nodata \\      
10-9-2002 & 52556 & F & 183 & \nodata & \nodata \\ 
11-6-2002 & 52585 & F & 174 & \nodata & \nodata \\
8-28-2003 & 52880 & F & 153 & \nodata & \nodata \\
8-29-2003 & 52881 & F & 166 & \nodata & \nodata \\     
8-31-2003 & 52883 & F & 159 & \nodata & \nodata \\
9-3-2003 &   52886 & P & 133  & 2.91 $\pm$ 0.01\% & 174 \\
9-25-2003 & 52908 & F & 162 & \nodata & \nodata \\ 
9-29-2003 & 52912 & F & 160 & \nodata & \nodata \\
10-13-2003 & 52926 & P & 137  & 2.73 $\pm$ 0.01\% & 0 \\
6-3-2004 &   53160 & P & 119  & 2.32 $\pm$ 0.02\% & 173 \\
8-13-2004 & 53231 & F & 161 & \nodata & \nodata \\
8-30-2004 &  53248 & P & 115  & 2.98 $\pm$ 0.02\% & 175 \\
9-12-2004 & 53261 & F & 148 & \nodata & \nodata \\
9-13-2004 & 53262 & F & 154 & \nodata & \nodata \\
9-27-2004 &  53276 & P & 121  & 2.88 $\pm$ 0.03\% & 179 \\
10-8-2004 & 53288 & F & 151 & \nodata & \nodata \\
10-11-2004 & 53290 & F & 148 & \nodata & \nodata \\
10-14-2004 & 53293 & F & 146 & \nodata & \nodata \\

\tablecomments{Summary of MWC 314 observations.  The date is tabulated
according to MM-DD-YYYY and MJD is an abbreviation for the modified Julian
date of the observation.  Data obtained at Pine Bluff Observatory is
labeled with a ``P'' in column 3, while data obtained at the Fesenkov
Astrophysical Institute is labeled with a ``F'' in column 3.  
Also listed is the H$\alpha$ equivalent width
(EW), with positive values corresponding to emission, the raw 
polarization in the R filter, and the raw position angle
(PA) in the R filter.  As described in the text, our FAI EWs appear to be systematically 
$\sim$28\AA larger than those derived from our PBO data.} 

\end{tabular}
\end{table}
\clearpage

\end{document}